\documentclass[pra,twocolumn,showpacs,amsmath,amssymb,superscriptaddress,longbibliography,reprint]{revtex4-1}
\usepackage{graphicx,bm,color,mathptmx} 

\usepackage[colorlinks=true,linkcolor=blue,citecolor=blue,urlcolor =blue]{hyperref}

\newcommand{\op}[1]{\hat{#1}}
\newcommand{\Tr}{\mathop{\mathrm{Tr}}\nolimits}

\begin{document}

\title{Local sampling of the Wigner function at telecom wavelength \\
with loss-tolerant detection of photon statistics}

\author{G.~Harder} 
\affiliation{Department of Physics, University of Paderborn, 
Warburger Stra{\ss}e 100, 33098 Paderborn, Germany}

\author{Ch.~Silberhorn} 
\affiliation{Department of Physics, University of Paderborn,
 Warburger Stra{\ss}e 100, 33098 Paderborn, Germany}
\affiliation{Max-Planck-Institut f\"ur die Physik des Lichts,
  G\"{u}nther-Scharowsky-Stra{\ss}e 1, Bau 24, 91058 Erlangen,
  Germany}

\author{J.~Rehacek} 
\affiliation{Department of Optics, Palack\'{y}  University, 
17. listopadu 12, 77146 Olomouc, Czech Republic}

\author{Z.~Hradil}
\affiliation{Department of Optics, Palack\'{y}  University, 
17. listopadu 12, 77146 Olomouc, Czech Republic}

\author{L.~Motka} 
\affiliation{Department of Optics, Palack\'{y}  University, 
17. listopadu 12, 77146 Olomouc, Czech Republic}

\author{B.~Stoklasa} 
\affiliation{Department of Optics, Palack\'{y}  University, 
17. listopadu 12, 77146 Olomouc, Czech Republic}

\author{L.~L.~S\'{a}nchez-Soto} 
\affiliation{Departamento de \'Optica, Facultad de F\'{\i}sica, 
 Universidad Complutense, 28040 Madrid, Spain}
\affiliation{Max-Planck-Institut f\"ur  die Physik des Lichts, 
 G\"{u}nther-Scharowsky-Stra{\ss}e 1, Bau 24,  91058 Erlangen, 
Germany}

\begin{abstract}
  We report the experimental point-by-point sampling of the Wigner
  function for nonclassical states created in an ultrafast pulsed
  type-II parametric down-conversion source. We use a loss-tolerant
  time-multiplexed detector based on a fiber-optical setup and a pair
  of photon-number-resolving avalanche photodiodes. By capitalizing on
  an expedient data-pattern tomography, we assess the properties of
  the light states with outstanding accuracy.  The method allows us to
  reliably infer the squeezing of genuine two-mode states without any
  phase reference.
\end{abstract}


\maketitle

Modern quantum technologies hinge on the capability to generate,
manipulate, and measure quantum states. A successful experimental
procedure requires verification of each of these steps: this is
the scope of quantum tomography~\cite{lnp:2004uq}.

Light is of particular significance in most of those developments
because it is unique as an information carrier. In quantum optics, the
continuous-variable community is mostly concerned with the wave
aspects of light, looking chiefly at quantities such as quantum noise,
squeezing, and entanglement. Homodyne detection is then the technique
of choice since it allows for a direct reconstruction of the Wigner
function in terms of the field quadratures~\cite{Lvovsky:2009jk}.

Interestingly enough, the Wigner function can also be determined at a
single point in phase space by measuring the
parity~\cite{Royer:1985ix}. The complete function can then be sampled
if this measurement is performed at a sufficient number of
points. This direct probing was first demonstrated with motional
states of a trapped ion~\cite{Leibfried:1996df} and applied with great
success to Fock states in cavity quantum
electrodynamics~\cite{Bertet:2002lo}.

In optics, this method can be implemented with a highly asymmetric
beam splitter, which superimposes the quantum signal of interest with
a coherent reference field (which occasionally can be as weak as the
signal~\cite{Kuzmich:2000aa,Donati:2014aa}), followed by photon
counting~\cite{Wallentowitz:1996qo,Banaszek:1996ng}.  Despite its
simplicity, the approach places stringent demands on detector
performance and requires a full photon-number resolving (PNR)
capability~\cite{Hadfield:2009qr,OBrien:2009fk,Buller:2010cl,
Natarajan:2012bf,Calkins:2013sf}, which, in fact, limits its practical
applicability. The same complication arises in other alternative
methods~\cite{Wenger:2004aa}.

A promising PNR detection strategy has recently been
demonstrated~\cite{Achilles:2003cs,Rehacek:2003bh,Fitch:2003uq}.  It
is based on time-multiplexed detection (TMD) with avalanche
photodiodes (APDs) and works even with pulses~\cite{Achilles:2006fk}.
For visible light, this has been used to implement a first direct
probing for heralded single photons~\cite{Laiho:2010kl}.  However, at
the telecom wavelengths employed in our experiment, the established
technology is based on InGaAs APDs, which are plagued by high
dark-count rates and long dead times, thereby making gating essential.

The effective implementation of these advanced schemes thus demands an
accurate knowledge of the detector~\cite{Luis:1999yg,
  Fiurasek:2001dn,DAriano:2004oe,Lundeen:2009sf,Amri:2011fk,
  Zhang:2012fu,Brida:2012mz}.  To address this issue, we adopt here the
so-called fitting of data patterns, an approach proposed in
Refs.~\cite{Rehacek:2010fk,Mogilevtsev:2013kl} and experimentally
realized in several groups~\cite{Cooper:2014qf,
  Harder:2014tf,Altorio:2016aa}. This technique enables calibration of
detectors with a sizable number of outcomes and their subsequent use
in state estimation.  It does not extract the complete set of
operators that describe the detector, but rather uses the raw
measurement outcome distributions for known input states as the
detector calibration. Besides, there is no need for any numerical 
postprocessing. In this Letter we use the pattern tomography as an
efficient tool to probe point by point the Wigner function of pulsed,
single photons and two-mode squeezed states at telecom
wavelengths. Thereby, we achieve the inference of mode-selective,
genuine two-mode squeezing without phase reference.

For a single-mode field, represented by the density operator
$\op{\varrho}$, the direct probing is based on the observation that
the Wigner function at the point $\alpha$ of the phase space can be
expressed as the average value of the parity operator on the probed
quantum state $\op{\varrho}$ displaced by
$- \alpha$~\cite{Wallentowitz:1996qo,Banaszek:1996ng}: viz,
\begin{equation}
  W( \alpha ) = \frac{2}{\pi}  \Tr  \left ( {\op{D}^{\dagger} (\alpha) 
      \op{\varrho} \op{D}(\alpha ) \, \op{\Pi}} \right ) =  
  \frac{2}{\pi} \sum_{n} (-1)^{n} \,  \varrho_{nn} (- \alpha) \, .
\end{equation}
Here, $\op{\Pi} = \exp ( i \pi \op{a}^{\dagger} \op{a} )$ is the
parity, whose eigenstates are the Fock states $|n \rangle$, with
eigenvalues $(-1)^{n}$, $\op{a}$ and $\op{a}^\dagger$ are
photon annihilation and creation operators,
$\op{\varrho} (- \alpha) = \op{D}^{\dagger} (\alpha) \op{\varrho}
\op{D}(\alpha )$, and $\op{D}(\alpha) = \exp ( \alpha \op{a}^{\dagger}
-  \alpha^{\ast} \op{a})$ is the displacement. 

There are significant differences between this direct probing and
homodyne tomography. The latter is a Gaussian measurement that
projects the state along a field quadrature. A PNR detector, on the
other hand, projects onto the photon-number basis, which is a
non-Gaussian operation.

Another important distinction is the fact that homodyne detection
involves an intrinsic filtering: only the part of the signal that
overlaps with the local oscillator can be seen by the detector. Losses
and mode mismatch yield the same signature.

Conversely, direct probing detects all the modes, which results in a
more complete characterization with an intrinsic quantification of the
mode overlap. For a nonunity overlap $\mathcal{M}$ with the reference
beam, the measured Wigner function is the product of the state and
vacuum Wigner functions: $W(\alpha) = W_{\textrm{state}} 
(\sqrt{\mathcal{M}} \alpha) W_{\textrm{vac}} ( \sqrt{1- \mathcal{M}} \alpha )$.
This behavior is fundamentally different from losses and is crucial
to estimating both the losses and the overlap with the local oscillator.

To sum up, homodyne detection is mode selective and insensitive to
detrimental effects from other modes and background. Direct probing is
sensitive to all modes and to the spatial-spectral single-mode
characteristics of the states. 

To demonstrate the capabilities of the direct probing, we choose a
pulsed single-photon state with a single-mode spectral structure;
i.e., with a diagonal density matrix. This simplifies the detection
scheme, preventing the need for phase sensitivity.  Consequently, we
use parametric down-conversion (PDC), a nonlinear process in which one
pump photon decays into two twin photons, called the signal and
the idler. Ideally, the output of a PDC source is the two-mode squeezed
vacuum
\begin{equation}
  | \Psi \rangle = \sqrt{1-\lambda^2} 
  \sum_n \lambda^n |{n,n} \rangle \, , 
  \label{eq:sqvac}
\end{equation}
where $n$ is the photon number in each mode, $\lambda=\tanh( r)$, and
$r$ is the squeezing parameter, which scales linearly with the pump field
amplitude, the nonlinear coefficient $\chi^{(2)}$ of the medium, and
the interaction length inside the crystal.

In the absence of losses, the perfect photon-number correlations of this
state allow for heralding Fock states in one mode by conditioning on
certain photon numbers in the other. With losses, the heralded states
can still look pretty much like Fock states and can show nonclassical
statistics.

\begin{figure}
  \includegraphics[width=0.98\columnwidth]{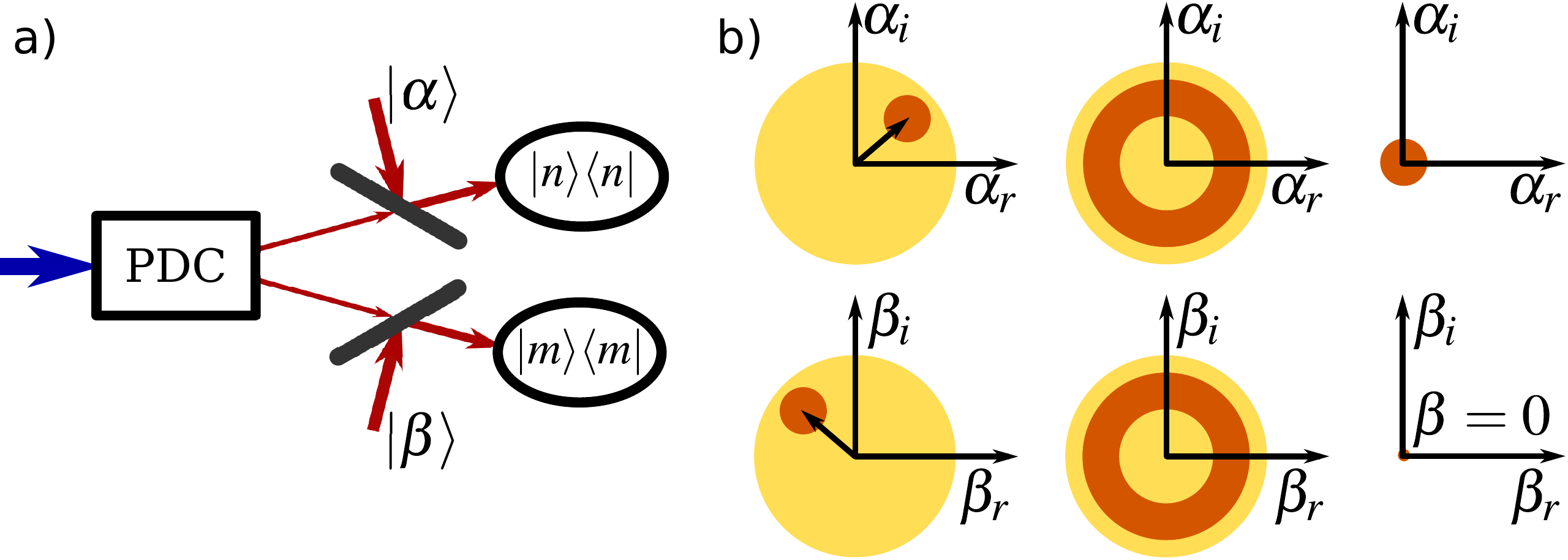}
  \caption{Phase-space correlations in PDC. (a) Direct
    probing scheme with two displacement beams, $\alpha$ and $\beta$,
    and PNR measurements. The marginal distributions of signal or
    idler are indicated by the larger circles, while correlations are
    indicated by the darker circles. (b) Phase-averaged state, where
    only the amplitudes remain correlated. Specifically, the
    distribution for $\beta=0$ shows correlations below the
    corresponding ones for the vacuum, from which the initial
    squeezing can be inferred.}
  \label{fig:scheme}
\end{figure}

These correlations can be clearly appreciated using the Wigner
function of the PDC state \eqref{eq:sqvac}, which is
\begin{equation}
  W (\alpha,\beta) = \frac{4}{\pi^2} 
  \exp \left ( 
    -e^{2r} |\alpha + \beta^{\ast} e^{-i\phi}|^2 - 
    e^{-2r} |\alpha - \beta^*e^{-i\phi}|^2
  \right ) \, , 
\end{equation}
where $\alpha$ and $\beta$ are the complex quadratures of the modes
and $\phi$ comes from the phase of the pump.  For infinite squeezing
$r\rightarrow \infty$, we have $|\alpha+\beta^*e^{-i\phi}| = 0$; i.e.,
the positions (real parts $\alpha_r$ and $\beta_r$) and the momenta
(imaginary parts $\alpha_i$ and $\beta_i$) are perfectly correlated or
anticorrelated, as indicated by the arrows in
Fig.~\ref{fig:scheme}(a). For finite $r$, the variance of the
correlations is $\mathrm{Var} (|\alpha+\beta^*e^{-i\phi}|) = e^{-2r}$,
which is below the Heisenberg limit of coherent states by a factor of
$e^{-2r}$, a signature of entanglement between the two modes. Such
correlations could be measured in a balanced homodyne setup, but it is
experimentally challenging, as pump and local oscillator phases have
to be locked.

Since we cannot access this phase, this is tantamount
to averaging over it, which yields the state
\begin{eqnarray}
  W_{\mathrm{avg}} (\alpha, \beta) & = & 
 \frac{4}{\pi^2} \exp[ -2\cosh (2r) (|\alpha|^2 + |\beta|^2)] \nonumber \\
& \times & I_0 \left ( -4 \sinh(2r) |\alpha||\beta| \right ) \, , 
\end{eqnarray} 
where $I_0(x)$ is the modified Bessel function. This
$W_{\mathrm{avg}}$ depends only on the amplitudes $|\alpha|$ and
$|\beta|$. Correspondingly, one can only measure photon numbers and
obtain information about the diagonal elements of the density
matrix. The off-diagonal elements remain hidden and entanglement is
lost.

Nonetheless, the correlations are still present around the origin, as
sketched in Fig.~\ref{fig:scheme}(b): if we set $\beta=0$, the
distribution of $|\alpha|$ is narrower by a factor of $\cosh(2r)$ than
the corresponding one for the vacuum, which can be verified with
direct probing.

\begin{figure}
  \includegraphics[width=0.84\columnwidth]{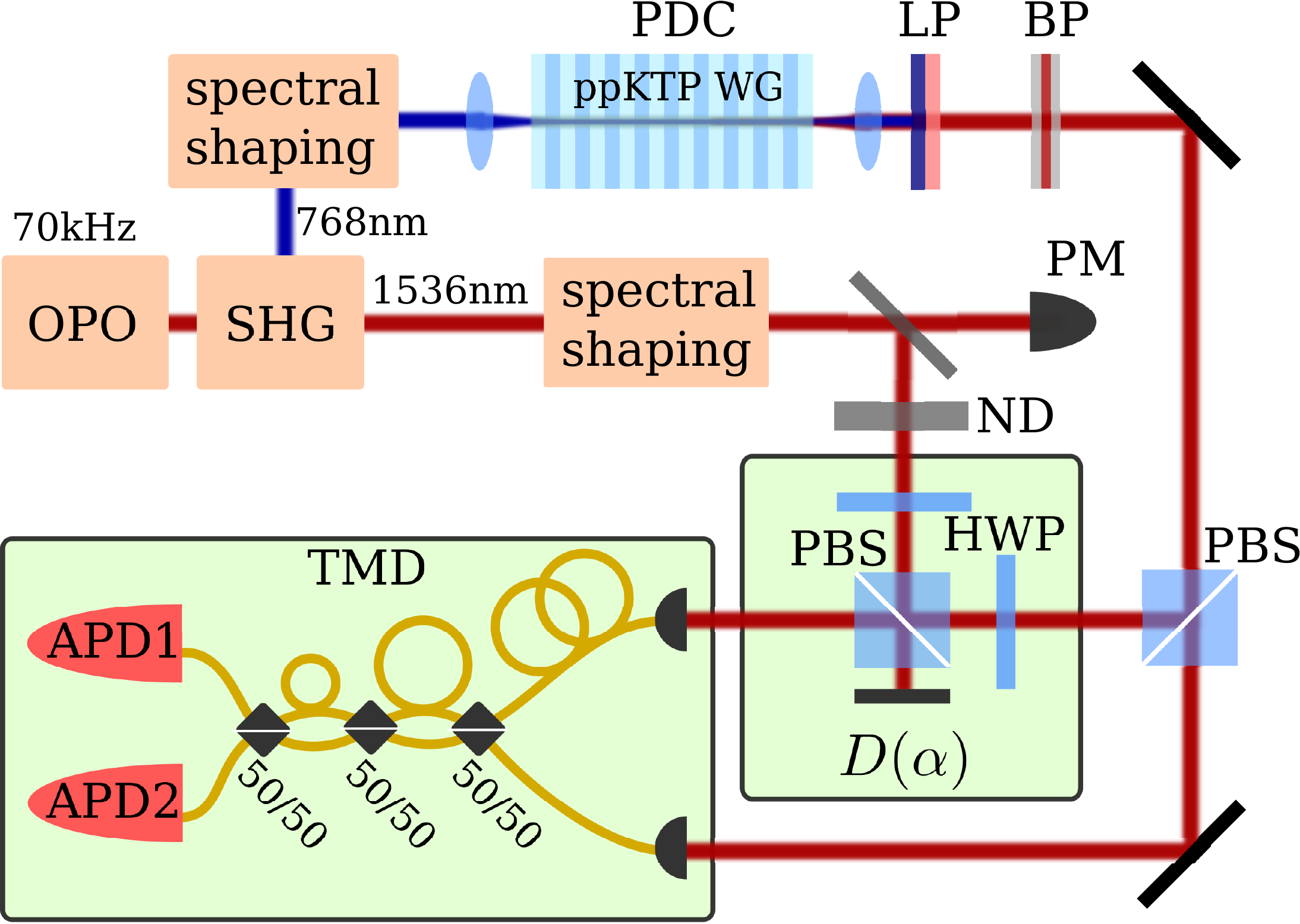}
  \caption{Experimental setup. Pulsed light from an
    optical parametric oscillator (OPO) is frequency doubled by second
    harmonic generation (SHG) and spectrally shaped in a
    4$f$-line. After the PDC, the pump is filtered with a long pass
    filter (LP) and the sinc sidelopes of the PDC spectrum and
    unwanted background are suppressed with a bandpass filter
    (BF). Signal and idler are split at a polarizing beam splitter
    (PBS).The reference beam, monitored with a power meter (PM), is
    filtered to the single-photon level using neutral density filters
    (ND) and overlapped with the idler at a 90:10 coupler consisting
    of a half wave plate (HWP) and a PBS. This realizes the
    displacement operation $\op{D} (\alpha)$.  Both states are coupled
    into single mode fibers and measured with an eight-bin, two-mode
    TMD.}
  \label{fig:setup}
\end{figure}

Our experimental setup is schematized in Fig.~\ref{fig:setup}. We use
PDC in an 8~mm long periodically poled KTP waveguide.  The process is
pumped with 0.5~ps pulses at a wavelength of 768~nm, producing
orthogonally polarized signal and idler beams at 1536~nm. The signal
and idler modes are decorrelated in frequency, enabling the heralding
of nearly pure states~\cite{Harder:2013iq}.

We overlap the signal with a coherent reference beam in an asymmetric
beam splitter with a 90:10 splitting ratio, consisting of a half-wave
plate and a polarizing beam splitter. In the limit of unity
transmission and perfect mode overlap this corresponds to the
displacement operator $\op{D}(\alpha)$~\cite{Wallentowitz:1996qo}. To
maximize the mode overlap, the reference beam has to match the signal
mode, both spectrally and temporally. Therefore, we perform spectral
shaping with a 4$f$ spectrometer made of two gratings and two lenses,
all separated by the focal length of the lenses. In the Fourier plane
of this spectrometer we use a variable slit which can be rotated
around the propagation axis: this allows us to shape the reference
spectrum to a Gaussian form matching the spectrum of the PDC mode. The
temporal overlap is achieved with a translation stage in the reference
beam. From Hong-Ou-Mandel dips between the reference and the signal we
estimate the overlap $\mathcal{M}$ to be around $0.7$.

Both output ports are measured with a TMD. One mode is delayed at the
input of the TMD and then each mode is split into four temporal bins
and two spacial bins at 50:50 beam splitters. In total, this amounts
to 16 bins such that up to eight photons per mode can be
measured. Each pulse impinges onto one of two InGaAs APDs with
detection efficiencies around~$20\%$. InGaAs APDs have the detrimental
effect of afterpulsing, which means that after a detection there is a
finite probability ($\sim7\%$) of a subsequent detection without a
photon being present. This makes the analysis of TMD data extremely
difficult, as afterpulses modify the photon-number distribution in a
nontrivial way. 

Some detector tomography is therefore necessary to reliably extract
the photon statistics of the incoming modes. Fitting of data patterns is
ideally suited for this task (see Supplemental Material for
details~\cite{suppl}). It requires the knowledge of the detector outputs for
certified input probes. We use coherent states already present as the
reference beam and couple them into both inputs of the TMD with
different power settings adjusted by motorized half-wave plates and
polarizing beam splitters.

\begin{figure}
  \centerline{\includegraphics[width=0.85\columnwidth]{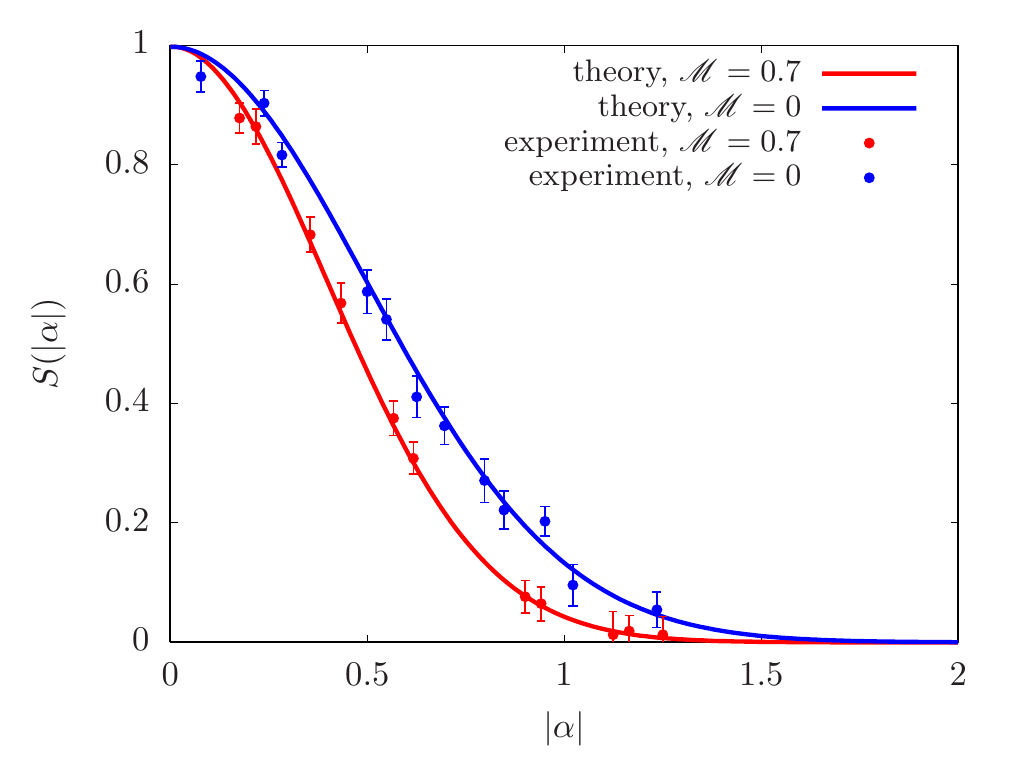}}
  \caption{Experimentally determined parities of a
    two-mode displaced PDC state with $r\approx 0.6$ as obtained with
    high overlap ($\mathcal{M}=0.7$, red symbols) and no overlap
    ($\mathcal{M}=0$, blue symbols) between the reference and signal
    modes. Theoretical predictions are also shown (solid lines).}
  \label{fig:YESvsNO}
\end{figure}

Powers are calibrated with respect to the power at the input fiber of
the TMD. Hence, losses inside the TMD and quantum efficiency of the
APDs are accounted for by the tomography. Losses inside the waveguide,
subsequent optical elements, and the fiber incoupling must be included
in the reconstructed state. This is the most meaningful separation
between the generation and the detection parts, as it is the point at
which the source would be combined with other sources or integrated
into a larger network.

The displaced PDC states are measured with
the TMD. The joint photon-number distribution $P_{mn}$ of each
displaced state is estimated using pattern tomography and the
corresponding parity $S(|\alpha|)=\sum_{mn}(-1)^{m+n}
P_{mn}(|\alpha|)$  calculated, where $P_{mn}(|\alpha|)$ is the 
reconstructed two-mode photon-number distribution with signal 
mode displaced by $|\alpha|$.  The measured parity values can be 
immediately converted to a Wigner function: $W(|\alpha|)=4 
S(|\alpha|)/\pi^2$.

Pattern tomography makes use of a set of 639 two-mode coherent probes
with known signal/idler amplitudes in the range
$0<|\alpha(\xi)|<3.5$. The probes are measured with the TMD and the
corresponding patterns are recorded. A sample of 100 patterns is used
for any single reconstruction, as this matches well a typical number
of linearly independent TMD outcomes for moderately intense
signals. The reconstruction is repeated with different randomly chosen
samples of pattern subsets. This makes it possible to estimate the errors
and calculate the final reconstruction averaging over pattern
sampling.

\begin{figure}
  \centerline{\includegraphics[width=0.88\columnwidth]{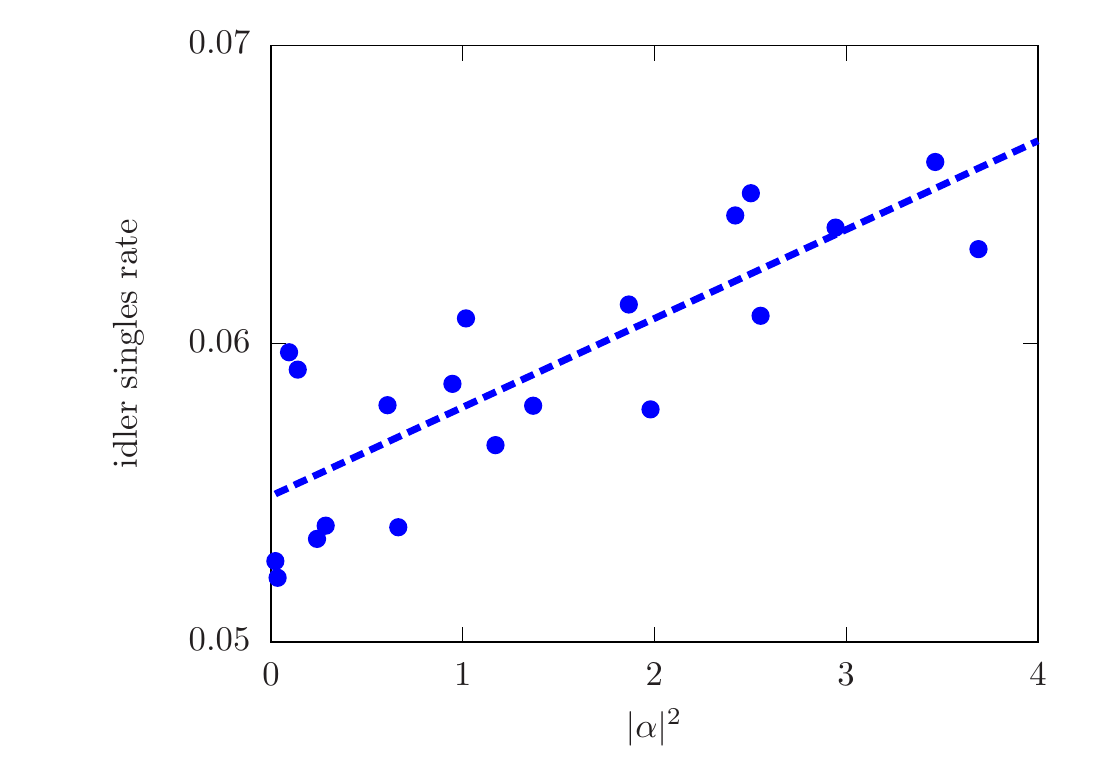}}
  \caption{Observed idler singles rate (dots) for different signal
    displacements of a PDC state with $r\approx 0.6$. Best linear fit
    is also shown (line).}
  \label{fig:Afterpulses}
\end{figure}

To compare the experimentally obtained parities with the theoretical
predictions, one has to know the reference displacements, as well as
the squeezing $r$ and the coupling efficiency $\eta$.  Displacements
are determined by measuring the reference beam alone and comparing the
probabilities of zero detection with those of the coherent probes. A
least-squares fit is sufficient for this purpose. Additionally, the
original undisplaced PDC state is measured to estimate the squeezing
parameter $r \approx 0.6$ and coupling efficiency $\eta \approx 0.75$:
these are the values for which the theory provides the best fit.

Figure~\ref{fig:YESvsNO} shows the parity of the displaced PDC state,
as estimated from the TMD data, for various displacements. For
moderate $|\alpha|$, the measured parity is below the corresponding
one for the vacuum and the Wigner function becomes slightly narrower,
witnessing the presence of the quadrature squeezing. This effect is
small, given the limited amount of squeezing available, but
statistically significant and also happens with no mode overlap. We
emphasize that, although this narrowing has been discussed already
for pure states, as in Eq.~\eqref{eq:sqvac}, it also reveals squeezing
for mixed states.

Next, we focus on heralding single-mode states by a single idler
click. Ideally, such a heralding generates a single photon in the signal
mode, a state characterized by a strongly negative Wigner function at
the origin.

Heralding at telecom wavelengths is highly nontrivial due to the
afterpulses in the idler counts. This effect can be ignored for weak
undisplaced PDC states~\cite{Harder:2014tf}, but, for large
displacements, afterpulses make up a significant fraction of the small
number of genuine idler detections (more than $10\%$ in our
case). This is illustrated in Fig.~\ref{fig:Afterpulses}, where the
idler single-click detection rate is plotted as a function of the
reference beam intensity $|\alpha|^{2}$.  Without afterpulsing, the
idler rates should stay constant, while we observe a linear increase
of the idler detection rate with $|\alpha|^{2}$.

\begin{figure}
  \centerline{\includegraphics[width=0.90\columnwidth]{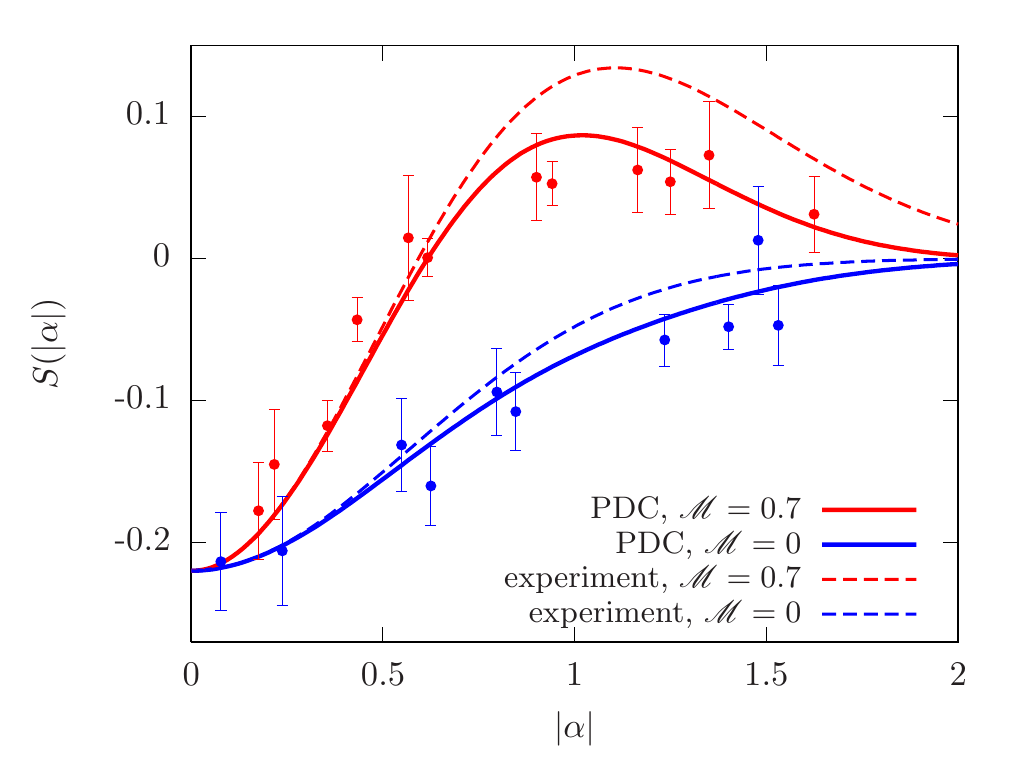}}
  \caption{Parities of heralded signal states measured
    with high overlap (red symbols) and no overlap (blue symbols) and
    $r\approx 0.6$. Solid (broken) lines show theory with afterpulses
    included (ignored). Notice that the afterpulses tend
    to decrease the measured parity of the displaced signal mode.}
  \label{fig:DispHeralded}
\end{figure}

After a single idler detection, the postmeasurement state of the
signal mode is $P_{s} = \Tr_{i} ( \op{E} \, \op{P} \,
\op{E}^{\dagger})/  \Tr_{s,i} (\op{E} \, \op{P} \, \op{E}^{\dagger})$,
where $\op{E}^\dag \op{E}= \op{\Pi}_{1s}$ is the operator describing
the idler detection, and $\Tr_{s,i}$ indicates tracing over both the
signal and idler modes. To deal with afterpulses, we construct the
measurement operator as an incoherent superposition of signal and
idler single detections, $\op{E}^\dag \op{E}=[1-x(\alpha)]
\op{\Pi}_{1i} +x (\alpha) \op{\Pi}_{1s}$.  The contribution
of afterpulses for a given displacement is estimated from 
Fig.~\ref{fig:Afterpulses}.

\begin{figure}
  \centerline{
    \includegraphics[width=\columnwidth]{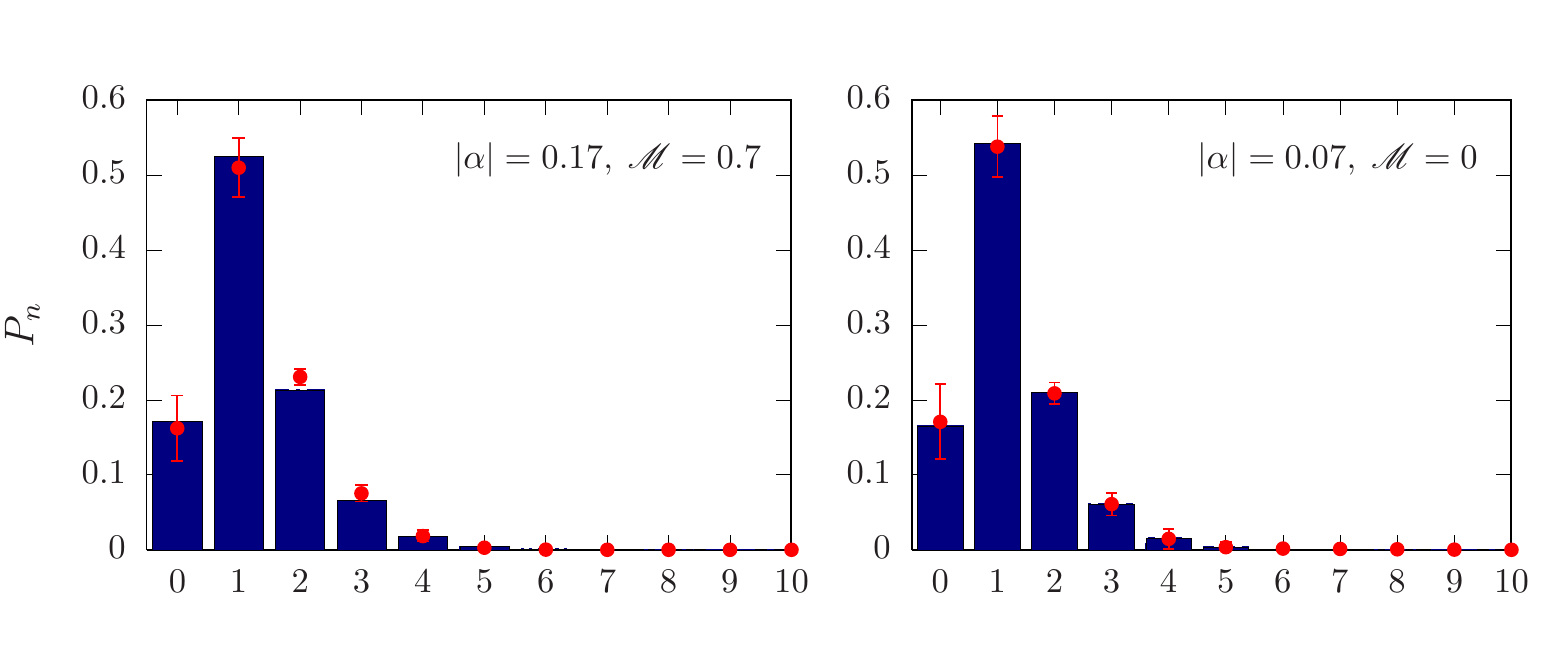}}
  \vspace{-7mm} \centerline{
    \includegraphics[width=\columnwidth]{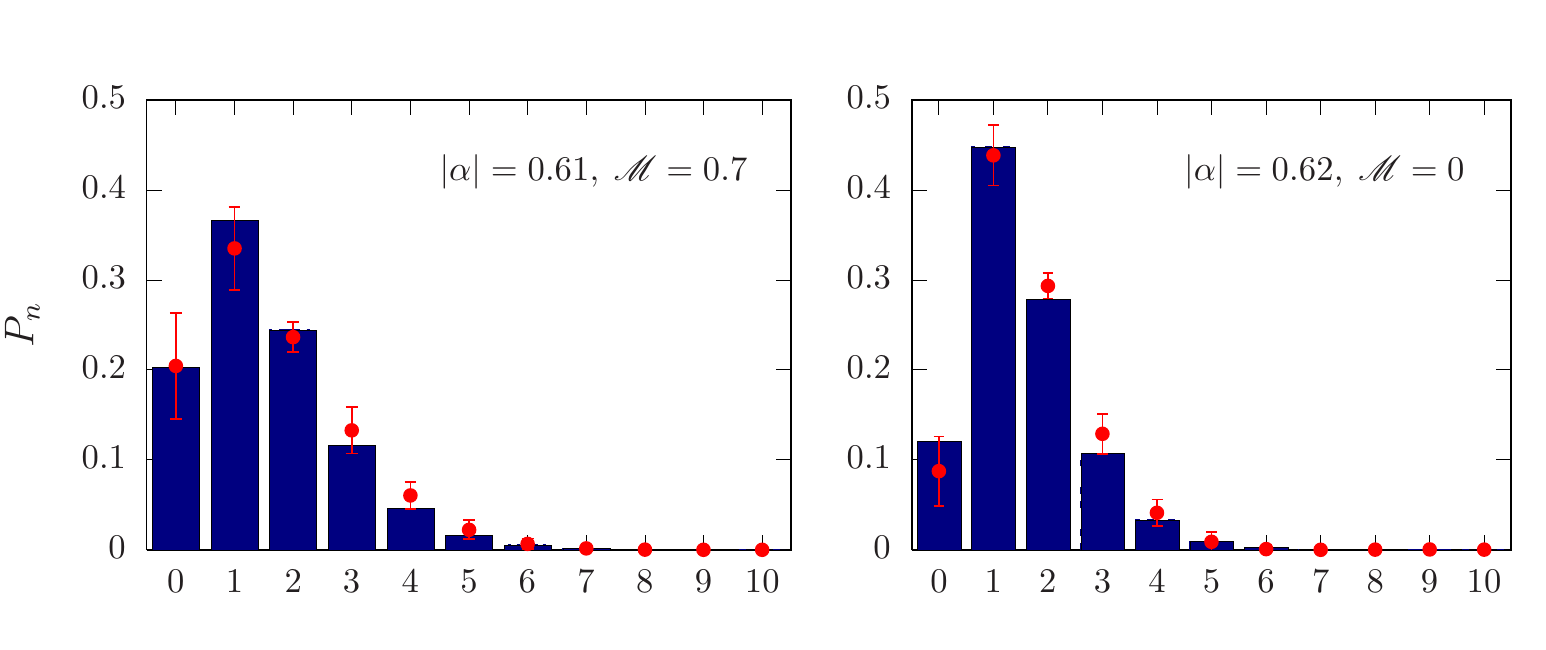}}
  \vspace{-7mm} \centerline{
    \includegraphics[width=\columnwidth]{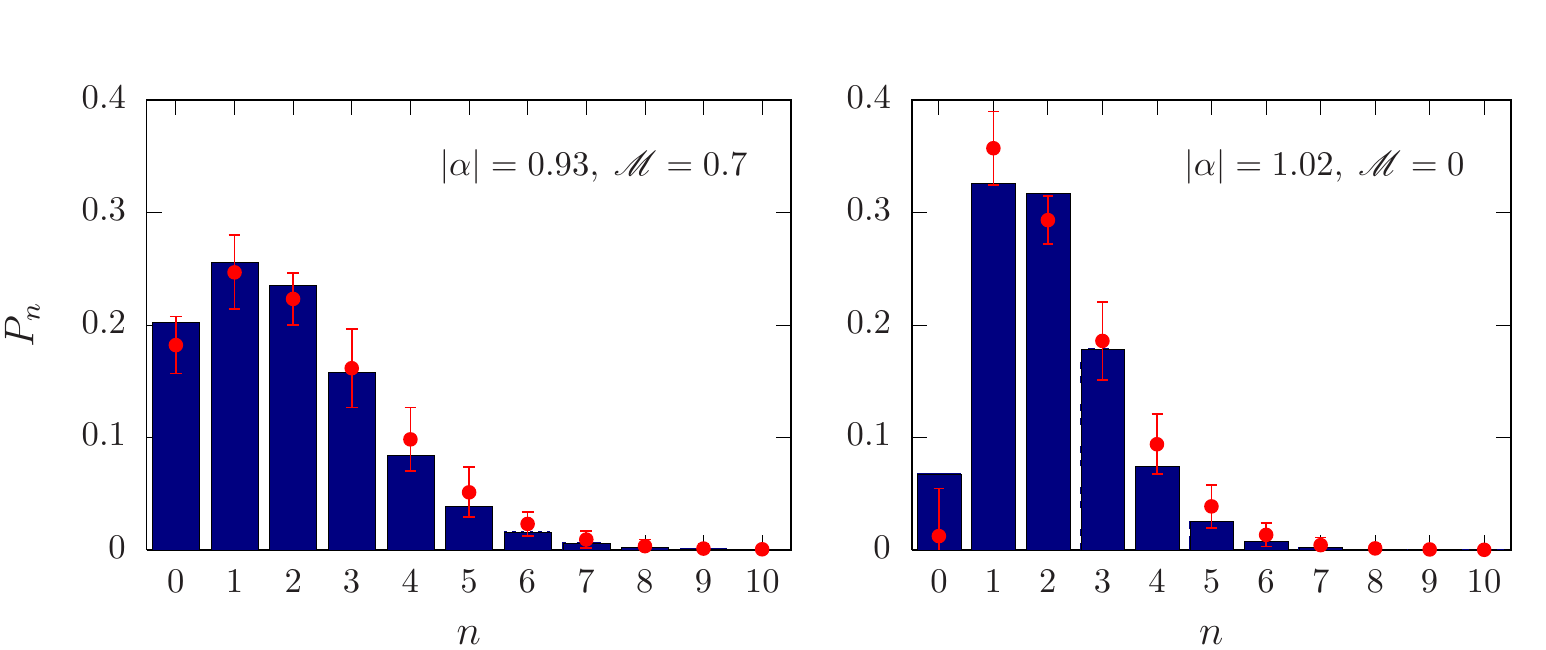}}
  \caption{Reconstructed signal photon-number
    distributions of heralded and displaced PDC states (dots);
    $r \approx 0.6$.  Theoretical models are also shown
    (bars). Displacement increases from top to bottom. Left (right)
    panels correspond to strong (no) overlap between the reference and
    signal modes.}
\label{fig:PnHeralded}
\end{figure}

Afterpulses introduce a systematic negative bias into the
reconstructed values of the signal parity. This is because heralding
by an afterpulse increases the single-photon component of the signal
state.  

Experimental characterization of displaced heralded states measured
with and without overlap between the signal and reference modes is
summarized in Fig.~\ref{fig:DispHeralded}. For high overlap, we expect
the parity to trace the Wigner function of the heralded signal state,
which can be approximated by a single-photon state. The negativity
around the origin is apparent in the red curves (corrected for
afterpulses). For no overlap, the measured parities remain negative
for all displacements. The two cases can be discriminated, showing
that the heralded states are very sensitive to previous interactions
in the signal beam.

In Fig.~\ref{fig:PnHeralded} we show the reconstructed signal
photon-number distributions $P_n$ for three different displacements
and $r \approx 0.6$. As the reference amplitude is increased, the
vacuum components of the signal states are
strongly suppressed. This could be expected. The heralded signal is
approximately a single-photon state, $P^s_n\approx\delta_{1,n}$. With
mode overlap switched on, this state is displaced by $|\alpha|$; with
mode overlap switched off, the resulting photon-number distribution is
simply a discrete convolution of the reference Poissonian distribution
$P^{\,r}_n=|\alpha|^{2n}\exp(-|\alpha|^2)/n!$ and the signal
single-photon distribution $P^s_n$. Obviously, the appearance of a
Kronecker delta in such a convolution makes the reference Poissonian
distribution displaced by one towards higher photon numbers,
$P_n\approx P^{\,r}_{n-1}$ effectively erasing the vacuum component
$P_0\approx 0$. 

It happens often that the full distribution $P_{mn}$ is not
required for a specific purpose. For example, only a few elements of
$P_{mn}$ or a linear function of $P_{mn}$, such as parity, might be
enough. This is called partial tomography and finds applications in
experiments with complex, highly dimensional systems, where a full
tomography is impractical or even impossible. A nice feature of the
pattern approach is that both full and partial tomography are done in
much the same way. A demonstration of the technique is presented in
the Supplemental Material.

In summary, we have directly probed the Wigner function of a
nonclassical single-photon wave packet in a robust, loss-tolerant
manner.  Our TMD detector, with the tool of data pattern, can verify
the nonclassicality and highlight the role of mode
properties in the detection.  This leads us to recognize the types of
experimental imperfections, and gives us valuable information about the
degradation caused by each one. 

G. H. and Ch. S. acknowledge the support from the EU Horizon 2020
Research and Innovation Program (Qcumber, Grant No 665148).  J. R.,
Z. H., L. M., and B. S. are thankful for the financial assistance of the Grant
Agency of the Czech Republic (Grant No 15-03194S), the IGA Project of
the Palack\'y University (Grant No. PRF 2016-005), and the Technology
Agency of the Czech Republic (Grant No TE01020229).

\newpage


\begin{thebibliography}{34}%
\makeatletter
\providecommand \@ifxundefined [1]{%
 \@ifx{#1\undefined}
}%
\providecommand \@ifnum [1]{%
 \ifnum #1\expandafter \@firstoftwo
 \else \expandafter \@secondoftwo
 \fi
}%
\providecommand \@ifx [1]{%
 \ifx #1\expandafter \@firstoftwo
 \else \expandafter \@secondoftwo
 \fi
}%
\providecommand \natexlab [1]{#1}%
\providecommand \enquote  [1]{``#1''}%
\providecommand \bibnamefont  [1]{#1}%
\providecommand \bibfnamefont [1]{#1}%
\providecommand \citenamefont [1]{#1}%
\providecommand \href@noop [0]{\@secondoftwo}%
\providecommand \href [0]{\begingroup \@sanitize@url \@href}%
\providecommand \@href[1]{\@@startlink{#1}\@@href}%
\providecommand \@@href[1]{\endgroup#1\@@endlink}%
\providecommand \@sanitize@url [0]{\catcode `\\12\catcode `\$12\catcode
  `\&12\catcode `\#12\catcode `\^12\catcode `\_12\catcode `\%12\relax}%
\providecommand \@@startlink[1]{}%
\providecommand \@@endlink[0]{}%
\providecommand \url  [0]{\begingroup\@sanitize@url \@url }%
\providecommand \@url [1]{\endgroup\@href {#1}{\urlprefix }}%
\providecommand \urlprefix  [0]{URL }%
\providecommand \Eprint [0]{\href }%
\providecommand \doibase [0]{http://dx.doi.org/}%
\providecommand \selectlanguage [0]{\@gobble}%
\providecommand \bibinfo  [0]{\@secondoftwo}%
\providecommand \bibfield  [0]{\@secondoftwo}%
\providecommand \translation [1]{[#1]}%
\providecommand \BibitemOpen [0]{}%
\providecommand \bibitemStop [0]{}%
\providecommand \bibitemNoStop [0]{.\EOS\space}%
\providecommand \EOS [0]{\spacefactor3000\relax}%
\providecommand \BibitemShut  [1]{\csname bibitem#1\endcsname}%
\let\auto@bib@innerbib\@empty
\bibitem [{\citenamefont {Paris}\ and\ \citenamefont
  {\v{R}eh\'a\v{c}ek}(2004)}]{lnp:2004uq}%
  \BibitemOpen
  \bibinfo {editor} {\bibfnamefont {M.~G.~A.}\ \bibnamefont {Paris}}\ and\
  \bibinfo {editor} {\bibfnamefont {J.}~\bibnamefont {\v{R}eh\'a\v{c}ek}},\
  eds.,\ \href@noop {} {\emph {\bibinfo {title} {Quantum State Estimation}}},\
  \bibinfo {series} {Lect. Not. Phys.}, Vol.\ \bibinfo {volume} {649}\
  (\bibinfo  {publisher} {Springer},\ \bibinfo {address} {Berlin},\ \bibinfo
  {year} {2004})\BibitemShut {NoStop}%
\bibitem [{\citenamefont {Lvovsky}\ and\ \citenamefont
  {Raymer}(2009)}]{Lvovsky:2009jk}%
  \BibitemOpen
  \bibfield  {author} {\bibinfo {author} {\bibfnamefont {A.~I.}\ \bibnamefont
  {Lvovsky}}\ and\ \bibinfo {author} {\bibfnamefont {M.~G.}\ \bibnamefont
  {Raymer}},\ }\bibfield  {title} {\enquote {\bibinfo {title}
  {Continuous-variable optical quantum-state tomography},}\ }\href@noop {}
  {\bibfield  {journal} {\bibinfo  {journal} {Rev. Mod. Phys.}\ }\textbf
  {\bibinfo {volume} {81}},\ \bibinfo {pages} {299--332} (\bibinfo {year}
  {2009})}\BibitemShut {NoStop}%
\bibitem [{\citenamefont {Royer}(1985)}]{Royer:1985ix}%
  \BibitemOpen
  \bibfield  {author} {\bibinfo {author} {\bibfnamefont {A.}~\bibnamefont
  {Royer}},\ }\bibfield  {title} {\enquote {\bibinfo {title} {Measurement of
  the {W}igner function},}\ }\href@noop {} {\bibfield  {journal} {\bibinfo
  {journal} {Phys. Rev. Lett.}\ }\textbf {\bibinfo {volume} {55}},\ \bibinfo
  {pages} {2745--2748} (\bibinfo {year} {1985})}\BibitemShut {NoStop}%
\bibitem [{\citenamefont {Leibfried}\ \emph {et~al.}(1996)\citenamefont
  {Leibfried}, \citenamefont {Meekhof}, \citenamefont {King}, \citenamefont
  {Monroe}, \citenamefont {Itano},\ and\ \citenamefont
  {Wineland}}]{Leibfried:1996df}%
  \BibitemOpen
  \bibfield  {author} {\bibinfo {author} {\bibfnamefont {D.}~\bibnamefont
  {Leibfried}}, \bibinfo {author} {\bibfnamefont {D.~M.}\ \bibnamefont
  {Meekhof}}, \bibinfo {author} {\bibfnamefont {B.~E.}\ \bibnamefont {King}},
  \bibinfo {author} {\bibfnamefont {C.}~\bibnamefont {Monroe}}, \bibinfo
  {author} {\bibfnamefont {W.~M.}\ \bibnamefont {Itano}}, \ and\ \bibinfo
  {author} {\bibfnamefont {D.~J.}\ \bibnamefont {Wineland}},\ }\bibfield
  {title} {\enquote {\bibinfo {title} {Experimental determination of the
  motional quantum state of a trapped atom},}\ }\href@noop {} {\bibfield
  {journal} {\bibinfo  {journal} {Phys. Rev. Lett.}\ }\textbf {\bibinfo
  {volume} {77}},\ \bibinfo {pages} {4281--4285} (\bibinfo {year}
  {1996})}\BibitemShut {NoStop}%
\bibitem [{\citenamefont {Bertet}\ \emph {et~al.}(2002)\citenamefont {Bertet},
  \citenamefont {Auffeves}, \citenamefont {Maioli}, \citenamefont {Osnaghi},
  \citenamefont {Meunier}, \citenamefont {Brune}, \citenamefont {Raimond},\
  and\ \citenamefont {Haroche}}]{Bertet:2002lo}%
  \BibitemOpen
  \bibfield  {author} {\bibinfo {author} {\bibfnamefont {P.}~\bibnamefont
  {Bertet}}, \bibinfo {author} {\bibfnamefont {A.}~\bibnamefont {Auffeves}},
  \bibinfo {author} {\bibfnamefont {P.}~\bibnamefont {Maioli}}, \bibinfo
  {author} {\bibfnamefont {S.}~\bibnamefont {Osnaghi}}, \bibinfo {author}
  {\bibfnamefont {T.}~\bibnamefont {Meunier}}, \bibinfo {author} {\bibfnamefont
  {M.}~\bibnamefont {Brune}}, \bibinfo {author} {\bibfnamefont {J.~M.}\
  \bibnamefont {Raimond}}, \ and\ \bibinfo {author} {\bibfnamefont
  {S.}~\bibnamefont {Haroche}},\ }\bibfield  {title} {\enquote {\bibinfo
  {title} {Direct measurement of the {W}igner function of a one-photon {F}ock
  state in a cavity},}\ }\href@noop {} {\bibfield  {journal} {\bibinfo
  {journal} {Phys. Rev. Lett.}\ }\textbf {\bibinfo {volume} {89}},\ \bibinfo
  {pages} {200402} (\bibinfo {year} {2002})}\BibitemShut {NoStop}%
\bibitem [{\citenamefont {Kuzmich}\ \emph {et~al.}(2000)\citenamefont
  {Kuzmich}, \citenamefont {Walmsley},\ and\ \citenamefont
  {Mandel}}]{Kuzmich:2000aa}%
  \BibitemOpen
  \bibfield  {author} {\bibinfo {author} {\bibfnamefont {A.}~\bibnamefont
  {Kuzmich}}, \bibinfo {author} {\bibfnamefont {I.~A.}\ \bibnamefont
  {Walmsley}}, \ and\ \bibinfo {author} {\bibfnamefont {L.}~\bibnamefont
  {Mandel}},\ }\bibfield  {title} {\enquote {\bibinfo {title} {Violation of
  {B}ell's inequality by a generalized {E}instein-{P}odolsky-{R}osen state
  using homodyne detection},}\ }\href@noop {} {\bibfield  {journal} {\bibinfo
  {journal} {Phys. Rev. Lett.}\ }\textbf {\bibinfo {volume} {85}},\ \bibinfo
  {pages} {1349--1353} (\bibinfo {year} {2000})}\BibitemShut {NoStop}%
\bibitem [{\citenamefont {Donati}\ \emph {et~al.}(2014)\citenamefont {Donati},
  \citenamefont {Bartley}, \citenamefont {Jin}, \citenamefont {Vidrighin},
  \citenamefont {Datta}, \citenamefont {Barbieri},\ and\ \citenamefont
  {Walmsley}}]{Donati:2014aa}%
  \BibitemOpen
  \bibfield  {author} {\bibinfo {author} {\bibfnamefont {G.}~\bibnamefont
  {Donati}}, \bibinfo {author} {\bibfnamefont {T.~J.}\ \bibnamefont {Bartley}},
  \bibinfo {author} {\bibfnamefont {X.-M.}\ \bibnamefont {Jin}}, \bibinfo
  {author} {\bibfnamefont {M.-D.}\ \bibnamefont {Vidrighin}}, \bibinfo {author}
  {\bibfnamefont {A.}~\bibnamefont {Datta}}, \bibinfo {author} {\bibfnamefont
  {M.}~\bibnamefont {Barbieri}}, \ and\ \bibinfo {author} {\bibfnamefont
  {I.~A.}\ \bibnamefont {Walmsley}},\ }\bibfield  {title} {\enquote {\bibinfo
  {title} {Observing optical coherence across {F}ock layers with weak-field
  homodyne detectors},}\ }\href@noop {} {\bibfield  {journal} {\bibinfo
  {journal} {Nat. Commun.}\ }\textbf {\bibinfo {volume} {5}},\ \bibinfo {pages}
  {6584} (\bibinfo {year} {2014})}\BibitemShut {NoStop}%
\bibitem [{\citenamefont {Wallentowitz}\ and\ \citenamefont
  {Vogel}(1996)}]{Wallentowitz:1996qo}%
  \BibitemOpen
  \bibfield  {author} {\bibinfo {author} {\bibfnamefont {S.}~\bibnamefont
  {Wallentowitz}}\ and\ \bibinfo {author} {\bibfnamefont {W.}~\bibnamefont
  {Vogel}},\ }\bibfield  {title} {\enquote {\bibinfo {title} {Unbalanced
  homodyning for quantum state measurements},}\ }\href@noop {} {\bibfield
  {journal} {\bibinfo  {journal} {Phys. Rev. A.}\ }\textbf {\bibinfo {volume}
  {53}},\ \bibinfo {pages} {4528--4533} (\bibinfo {year} {1996})}\BibitemShut
  {NoStop}%
\bibitem [{\citenamefont {Banaszek}\ and\ \citenamefont
  {W{\'o}dkiewicz}(1996)}]{Banaszek:1996ng}%
  \BibitemOpen
  \bibfield  {author} {\bibinfo {author} {\bibfnamefont {K.}~\bibnamefont
  {Banaszek}}\ and\ \bibinfo {author} {\bibfnamefont {K.}~\bibnamefont
  {W{\'o}dkiewicz}},\ }\bibfield  {title} {\enquote {\bibinfo {title} {Direct
  probing of quantum phase space by photon counting},}\ }\href@noop {}
  {\bibfield  {journal} {\bibinfo  {journal} {Phys. Rev. Lett.}\ }\textbf
  {\bibinfo {volume} {76}},\ \bibinfo {pages} {4344--4347} (\bibinfo {year}
  {1996})}\BibitemShut {NoStop}%
\bibitem [{\citenamefont {Hadfield}(2009)}]{Hadfield:2009qr}%
  \BibitemOpen
  \bibfield  {author} {\bibinfo {author} {\bibfnamefont {R.~H.}\ \bibnamefont
  {Hadfield}},\ }\bibfield  {title} {\enquote {\bibinfo {title} {Single-photon
  detectors for optical quantum information applications},}\ }\href@noop {}
  {\bibfield  {journal} {\bibinfo  {journal} {Nat. Photon,}\ }\textbf {\bibinfo
  {volume} {3}},\ \bibinfo {pages} {696--705} (\bibinfo {year}
  {2009})}\BibitemShut {NoStop}%
\bibitem [{\citenamefont {O'Brien}\ \emph {et~al.}(2009)\citenamefont
  {O'Brien}, \citenamefont {Furusawa},\ and\ \citenamefont
  {Vuckovic}}]{OBrien:2009fk}%
  \BibitemOpen
  \bibfield  {author} {\bibinfo {author} {\bibfnamefont {J.~L.}\ \bibnamefont
  {O'Brien}}, \bibinfo {author} {\bibfnamefont {A.}~\bibnamefont {Furusawa}}, \
  and\ \bibinfo {author} {\bibfnamefont {J.}~\bibnamefont {Vuckovic}},\
  }\bibfield  {title} {\enquote {\bibinfo {title} {Photonic quantum
  technologies},}\ }\href@noop {} {\bibfield  {journal} {\bibinfo  {journal}
  {Nat. Photon.}\ }\textbf {\bibinfo {volume} {3}},\ \bibinfo {pages}
  {687--695} (\bibinfo {year} {2009})}\BibitemShut {NoStop}%
\bibitem [{\citenamefont {Buller}\ and\ \citenamefont
  {Collins}(2010)}]{Buller:2010cl}%
  \BibitemOpen
  \bibfield  {author} {\bibinfo {author} {\bibfnamefont {G.~S.}\ \bibnamefont
  {Buller}}\ and\ \bibinfo {author} {\bibfnamefont {R.~J.}\ \bibnamefont
  {Collins}},\ }\bibfield  {title} {\enquote {\bibinfo {title} {Single-photon
  generation and detection},}\ }\href@noop {} {\bibfield  {journal} {\bibinfo
  {journal} {Meas. Sci. Technol.}\ }\textbf {\bibinfo {volume} {21}},\ \bibinfo
  {pages} {012002} (\bibinfo {year} {2010})}\BibitemShut {NoStop}%
\bibitem [{\citenamefont {Natarajan}\ \emph {et~al.}(2012)\citenamefont
  {Natarajan}, \citenamefont {Tanner},\ and\ \citenamefont
  {Hadfield}}]{Natarajan:2012bf}%
  \BibitemOpen
  \bibfield  {author} {\bibinfo {author} {\bibfnamefont {C.~M.}\ \bibnamefont
  {Natarajan}}, \bibinfo {author} {\bibfnamefont {M.~G.}\ \bibnamefont
  {Tanner}}, \ and\ \bibinfo {author} {\bibfnamefont {R.~H.}\ \bibnamefont
  {Hadfield}},\ }\bibfield  {title} {\enquote {\bibinfo {title}
  {Superconducting nanowire single-photon detectors: physics and
  applications},}\ }\href@noop {} {\bibfield  {journal} {\bibinfo  {journal}
  {Supercond. Sci. Technol.}\ }\textbf {\bibinfo {volume} {25}},\ \bibinfo
  {pages} {063001} (\bibinfo {year} {2012})}\BibitemShut {NoStop}%
\bibitem [{\citenamefont {Calkins}\ \emph {et~al.}(2013)\citenamefont
  {Calkins}, \citenamefont {Mennea}, \citenamefont {Lita}, \citenamefont
  {Metcalf}, \citenamefont {Kolthammer}, \citenamefont {Lamas-Linares},
  \citenamefont {Spring}, \citenamefont {Humphreys}, \citenamefont {Mirin},
  \citenamefont {Gates}, \citenamefont {Smith}, \citenamefont {Walmsley},
  \citenamefont {Gerrits},\ and\ \citenamefont {Nam}}]{Calkins:2013sf}%
  \BibitemOpen
  \bibfield  {author} {\bibinfo {author} {\bibfnamefont {B.}~\bibnamefont
  {Calkins}}, \bibinfo {author} {\bibfnamefont {P.~L.}\ \bibnamefont {Mennea}},
  \bibinfo {author} {\bibfnamefont {A.~E.}\ \bibnamefont {Lita}}, \bibinfo
  {author} {\bibfnamefont {B.~J.}\ \bibnamefont {Metcalf}}, \bibinfo {author}
  {\bibfnamefont {W.~S.}\ \bibnamefont {Kolthammer}}, \bibinfo {author}
  {\bibfnamefont {A.}~\bibnamefont {Lamas-Linares}}, \bibinfo {author}
  {\bibfnamefont {J.~B.}\ \bibnamefont {Spring}}, \bibinfo {author}
  {\bibfnamefont {P.~C.}\ \bibnamefont {Humphreys}}, \bibinfo {author}
  {\bibfnamefont {R.~P.}\ \bibnamefont {Mirin}}, \bibinfo {author}
  {\bibfnamefont {J.~C.}\ \bibnamefont {Gates}}, \bibinfo {author}
  {\bibfnamefont {P.~G.~R.}\ \bibnamefont {Smith}}, \bibinfo {author}
  {\bibfnamefont {I.~A.}\ \bibnamefont {Walmsley}}, \bibinfo {author}
  {\bibfnamefont {T.}~\bibnamefont {Gerrits}}, \ and\ \bibinfo {author}
  {\bibfnamefont {S.~W.}\ \bibnamefont {Nam}},\ }\bibfield  {title} {\enquote
  {\bibinfo {title} {High quantum-efficiency photon-number-resolving detector
  for photonic on-chip information processing,},}\ }\href@noop {} {\bibfield
  {journal} {\bibinfo  {journal} {Opt. Express}\ }\textbf {\bibinfo {volume}
  {21}},\ \bibinfo {pages} {22657--22670} (\bibinfo {year} {2013})}\BibitemShut
  {NoStop}%
\bibitem [{\citenamefont {Wenger}\ \emph {et~al.}(2004)\citenamefont {Wenger},
  \citenamefont {Fiur{\'a}{\v s}ek}, \citenamefont {Tualle-Brouri},
  \citenamefont {Cerf},\ and\ \citenamefont {Grangier}}]{Wenger:2004aa}%
  \BibitemOpen
  \bibfield  {author} {\bibinfo {author} {\bibfnamefont {J.}~\bibnamefont
  {Wenger}}, \bibinfo {author} {\bibfnamefont {J.}~\bibnamefont {Fiur{\'a}{\v
  s}ek}}, \bibinfo {author} {\bibfnamefont {R.}~\bibnamefont {Tualle-Brouri}},
  \bibinfo {author} {\bibfnamefont {N.~J.}\ \bibnamefont {Cerf}}, \ and\
  \bibinfo {author} {\bibfnamefont {P.}~\bibnamefont {Grangier}},\ }\bibfield
  {title} {\enquote {\bibinfo {title} {Pulsed squeezed vacuum measurements
  without homodyning},}\ }\href@noop {} {\bibfield  {journal} {\bibinfo
  {journal} {Phys. Rev. A}\ }\textbf {\bibinfo {volume} {70}},\ \bibinfo
  {pages} {053812} (\bibinfo {year} {2004})}\BibitemShut {NoStop}%
\bibitem [{\citenamefont {Achilles}\ \emph {et~al.}(2003)\citenamefont
  {Achilles}, \citenamefont {Silberhorn}, \citenamefont {\'{S}liwa},
  \citenamefont {Banaszek},\ and\ \citenamefont {Walmsley}}]{Achilles:2003cs}%
  \BibitemOpen
  \bibfield  {author} {\bibinfo {author} {\bibfnamefont {D.}~\bibnamefont
  {Achilles}}, \bibinfo {author} {\bibfnamefont {Ch.}\ \bibnamefont
  {Silberhorn}}, \bibinfo {author} {\bibfnamefont {C.}~\bibnamefont
  {\'{S}liwa}}, \bibinfo {author} {\bibfnamefont {K.}~\bibnamefont {Banaszek}},
  \ and\ \bibinfo {author} {\bibfnamefont {I.~A.}\ \bibnamefont {Walmsley}},\
  }\bibfield  {title} {\enquote {\bibinfo {title} {Fiber-assisted detection
  with photon number resolution},}\ }\href@noop {} {\bibfield  {journal}
  {\bibinfo  {journal} {Opt. Lett.}\ }\textbf {\bibinfo {volume} {28}},\
  \bibinfo {pages} {2387--2389} (\bibinfo {year} {2003})}\BibitemShut {NoStop}%
\bibitem [{\citenamefont {{\v R}eh{\'a}{\v c}ek}\ \emph
  {et~al.}(2003)\citenamefont {{\v R}eh{\'a}{\v c}ek}, \citenamefont {Hradil},
  \citenamefont {Haderka}, \citenamefont {Pe{\v r}ina},\ and\ \citenamefont
  {Hamar}}]{Rehacek:2003bh}%
  \BibitemOpen
  \bibfield  {author} {\bibinfo {author} {\bibfnamefont {J.}~\bibnamefont {{\v
  R}eh{\'a}{\v c}ek}}, \bibinfo {author} {\bibfnamefont {Z.}~\bibnamefont
  {Hradil}}, \bibinfo {author} {\bibfnamefont {O.}~\bibnamefont {Haderka}},
  \bibinfo {author} {\bibfnamefont {J.}~\bibnamefont {Pe{\v r}ina}}, \ and\
  \bibinfo {author} {\bibfnamefont {M.}~\bibnamefont {Hamar}},\ }\bibfield
  {title} {\enquote {\bibinfo {title} {Multiple-photon resolving fiber-loop
  detector},}\ }\href@noop {} {\bibfield  {journal} {\bibinfo  {journal} {Phys.
  Rev. A}\ }\textbf {\bibinfo {volume} {67}},\ \bibinfo {pages} {061801}
  (\bibinfo {year} {2003})}\BibitemShut {NoStop}%
\bibitem [{\citenamefont {Fitch}\ \emph {et~al.}(2003)\citenamefont {Fitch},
  \citenamefont {Jacobs}, \citenamefont {Pittman},\ and\ \citenamefont
  {Franson}}]{Fitch:2003uq}%
  \BibitemOpen
  \bibfield  {author} {\bibinfo {author} {\bibfnamefont {M.~J.}\ \bibnamefont
  {Fitch}}, \bibinfo {author} {\bibfnamefont {B.~C.}\ \bibnamefont {Jacobs}},
  \bibinfo {author} {\bibfnamefont {T.~B.}\ \bibnamefont {Pittman}}, \ and\
  \bibinfo {author} {\bibfnamefont {J.~D.}\ \bibnamefont {Franson}},\
  }\bibfield  {title} {\enquote {\bibinfo {title} {Photon-number resolution
  using time-multiplexed single-photon detectors},}\ }\href@noop {} {\bibfield
  {journal} {\bibinfo  {journal} {Phys. Rev. A}\ }\textbf {\bibinfo {volume}
  {68}},\ \bibinfo {pages} {043814} (\bibinfo {year} {2003})}\BibitemShut
  {NoStop}%
\bibitem [{\citenamefont {Achilles}\ \emph {et~al.}(2006)\citenamefont
  {Achilles}, \citenamefont {Silberhorn},\ and\ \citenamefont
  {Walmsley}}]{Achilles:2006fk}%
  \BibitemOpen
  \bibfield  {author} {\bibinfo {author} {\bibfnamefont {D.}~\bibnamefont
  {Achilles}}, \bibinfo {author} {\bibfnamefont {Ch.}\ \bibnamefont
  {Silberhorn}}, \ and\ \bibinfo {author} {\bibfnamefont {I.~A.}\ \bibnamefont
  {Walmsley}},\ }\bibfield  {title} {\enquote {\bibinfo {title} {Direct,
  loss-tolerant characterization of nonclassical photon statistics},}\
  }\href@noop {} {\bibfield  {journal} {\bibinfo  {journal} {Phys. Rev. Lett.}\
  }\textbf {\bibinfo {volume} {97}},\ \bibinfo {pages} {043602} (\bibinfo
  {year} {2006})}\BibitemShut {NoStop}%
\bibitem [{\citenamefont {Laiho}\ \emph {et~al.}(2010)\citenamefont {Laiho},
  \citenamefont {Cassemiro}, \citenamefont {Gross},\ and\ \citenamefont
  {Silberhorn}}]{Laiho:2010kl}%
  \BibitemOpen
  \bibfield  {author} {\bibinfo {author} {\bibfnamefont {K.}~\bibnamefont
  {Laiho}}, \bibinfo {author} {\bibfnamefont {K.~N.}\ \bibnamefont
  {Cassemiro}}, \bibinfo {author} {\bibfnamefont {D.}~\bibnamefont {Gross}}, \
  and\ \bibinfo {author} {\bibfnamefont {Ch.}\ \bibnamefont {Silberhorn}},\
  }\bibfield  {title} {\enquote {\bibinfo {title} {Probing the negative wigner
  function of a pulsed single photon point by point},}\ }\href@noop {}
  {\bibfield  {journal} {\bibinfo  {journal} {Phys. Rev. Lett.}\ }\textbf
  {\bibinfo {volume} {105}},\ \bibinfo {pages} {253603} (\bibinfo {year}
  {2010})}\BibitemShut {NoStop}%
\bibitem [{\citenamefont {Luis}\ and\ \citenamefont
  {S{\'a}nchez-Soto}(1999)}]{Luis:1999yg}%
  \BibitemOpen
  \bibfield  {author} {\bibinfo {author} {\bibfnamefont {A.}~\bibnamefont
  {Luis}}\ and\ \bibinfo {author} {\bibfnamefont {L.~L.}\ \bibnamefont
  {S{\'a}nchez-Soto}},\ }\bibfield  {title} {\enquote {\bibinfo {title}
  {Complete characterization of arbitrary quantum measurement processes},}\
  }\href@noop {} {\bibfield  {journal} {\bibinfo  {journal} {Phys. Rev. Lett.}\
  }\textbf {\bibinfo {volume} {83}},\ \bibinfo {pages} {3573--3576} (\bibinfo
  {year} {1999})}\BibitemShut {NoStop}%
\bibitem [{\citenamefont {Fiur{\'a}{\v s}ek}(2001)}]{Fiurasek:2001dn}%
  \BibitemOpen
  \bibfield  {author} {\bibinfo {author} {\bibfnamefont {J.}~\bibnamefont
  {Fiur{\'a}{\v s}ek}},\ }\bibfield  {title} {\enquote {\bibinfo {title}
  {Maximum-likelihood estimation of quantum measurement},}\ }\href@noop {}
  {\bibfield  {journal} {\bibinfo  {journal} {Phys. Rev. A}\ }\textbf {\bibinfo
  {volume} {64}},\ \bibinfo {pages} {024102} (\bibinfo {year}
  {2001})}\BibitemShut {NoStop}%
\bibitem [{\citenamefont {D'Ariano}\ \emph {et~al.}(2004)\citenamefont
  {D'Ariano}, \citenamefont {Maccone},\ and\ \citenamefont
  {Lo~Presti}}]{DAriano:2004oe}%
  \BibitemOpen
  \bibfield  {author} {\bibinfo {author} {\bibfnamefont {G.~M.}\ \bibnamefont
  {D'Ariano}}, \bibinfo {author} {\bibfnamefont {L.}~\bibnamefont {Maccone}}, \
  and\ \bibinfo {author} {\bibfnamefont {P.}~\bibnamefont {Lo~Presti}},\
  }\bibfield  {title} {\enquote {\bibinfo {title} {Quantum calibration of
  measurement instrumentation},}\ }\href@noop {} {\bibfield  {journal}
  {\bibinfo  {journal} {Phys. Rev. Lett.}\ }\textbf {\bibinfo {volume} {93}},\
  \bibinfo {pages} {250407} (\bibinfo {year} {2004})}\BibitemShut {NoStop}%
\bibitem [{\citenamefont {Lundeen}\ \emph {et~al.}(2009)\citenamefont
  {Lundeen}, \citenamefont {Feito}, \citenamefont {Coldenstrodt-Ronge},
  \citenamefont {Pregnell}, \citenamefont {Silberhorn}, \citenamefont {Ralph},
  \citenamefont {Eisert}, \citenamefont {Plenio},\ and\ \citenamefont
  {Walmsley}}]{Lundeen:2009sf}%
  \BibitemOpen
  \bibfield  {author} {\bibinfo {author} {\bibfnamefont {J.~S.}\ \bibnamefont
  {Lundeen}}, \bibinfo {author} {\bibfnamefont {A.}~\bibnamefont {Feito}},
  \bibinfo {author} {\bibfnamefont {H.}~\bibnamefont {Coldenstrodt-Ronge}},
  \bibinfo {author} {\bibfnamefont {K.~L.}\ \bibnamefont {Pregnell}}, \bibinfo
  {author} {\bibfnamefont {Ch.}\ \bibnamefont {Silberhorn}}, \bibinfo {author}
  {\bibfnamefont {T.~C.}\ \bibnamefont {Ralph}}, \bibinfo {author}
  {\bibfnamefont {J.}~\bibnamefont {Eisert}}, \bibinfo {author} {\bibfnamefont
  {M.~B.}\ \bibnamefont {Plenio}}, \ and\ \bibinfo {author} {\bibfnamefont
  {I.~A.}\ \bibnamefont {Walmsley}},\ }\bibfield  {title} {\enquote {\bibinfo
  {title} {Tomography of quantum detectors},}\ }\href@noop {} {\bibfield
  {journal} {\bibinfo  {journal} {Nat. Phys.}\ }\textbf {\bibinfo {volume}
  {5}},\ \bibinfo {pages} {27--30} (\bibinfo {year} {2009})}\BibitemShut
  {NoStop}%
\bibitem [{\citenamefont {Amri}\ \emph {et~al.}(2011)\citenamefont {Amri},
  \citenamefont {Laurat},\ and\ \citenamefont {Fabre}}]{Amri:2011fk}%
  \BibitemOpen
  \bibfield  {author} {\bibinfo {author} {\bibfnamefont {T.}~\bibnamefont
  {Amri}}, \bibinfo {author} {\bibfnamefont {J.}~\bibnamefont {Laurat}}, \ and\
  \bibinfo {author} {\bibfnamefont {C.}~\bibnamefont {Fabre}},\ }\bibfield
  {title} {\enquote {\bibinfo {title} {Characterizing quantum properties of a
  measurement apparatus: Insights from the retrodictive approach},}\
  }\href@noop {} {\bibfield  {journal} {\bibinfo  {journal} {Phys. Rev. Lett.}\
  }\textbf {\bibinfo {volume} {106}},\ \bibinfo {pages} {020502} (\bibinfo
  {year} {2011})}\BibitemShut {NoStop}%
\bibitem [{\citenamefont {Zhang}\ \emph {et~al.}(2012)\citenamefont {Zhang},
  \citenamefont {Datta}, \citenamefont {Coldenstrodt-Ronge}, \citenamefont
  {Jin}, \citenamefont {Eisert}, \citenamefont {Plenio},\ and\ \citenamefont
  {Walmsley}}]{Zhang:2012fu}%
  \BibitemOpen
  \bibfield  {author} {\bibinfo {author} {\bibfnamefont {L.}~\bibnamefont
  {Zhang}}, \bibinfo {author} {\bibfnamefont {A.}~\bibnamefont {Datta}},
  \bibinfo {author} {\bibfnamefont {H.~B.}\ \bibnamefont {Coldenstrodt-Ronge}},
  \bibinfo {author} {\bibfnamefont {X.-M.}\ \bibnamefont {Jin}}, \bibinfo
  {author} {\bibfnamefont {J.}~\bibnamefont {Eisert}}, \bibinfo {author}
  {\bibfnamefont {M.~B.}\ \bibnamefont {Plenio}}, \ and\ \bibinfo {author}
  {\bibfnamefont {I.~A.}\ \bibnamefont {Walmsley}},\ }\bibfield  {title}
  {\enquote {\bibinfo {title} {Recursive quantum detector tomography},}\
  }\href@noop {} {\bibfield  {journal} {\bibinfo  {journal} {New J. Phys.}\
  }\textbf {\bibinfo {volume} {14}},\ \bibinfo {pages} {115005} (\bibinfo
  {year} {2012})}\BibitemShut {NoStop}%
\bibitem [{\citenamefont {Brida}\ \emph {et~al.}(2012)\citenamefont {Brida},
  \citenamefont {Ciavarella}, \citenamefont {Degiovanni}, \citenamefont
  {Genovese}, \citenamefont {Lolli}, \citenamefont {Mingolla}, \citenamefont
  {Piacentini}, \citenamefont {Rajteri}, \citenamefont {Taralli},\ and\
  \citenamefont {Paris}}]{Brida:2012mz}%
  \BibitemOpen
  \bibfield  {author} {\bibinfo {author} {\bibfnamefont {G.}~\bibnamefont
  {Brida}}, \bibinfo {author} {\bibfnamefont {L.}~\bibnamefont {Ciavarella}},
  \bibinfo {author} {\bibfnamefont {I.~P.}\ \bibnamefont {Degiovanni}},
  \bibinfo {author} {\bibfnamefont {M.}~\bibnamefont {Genovese}}, \bibinfo
  {author} {\bibfnamefont {L.}~\bibnamefont {Lolli}}, \bibinfo {author}
  {\bibfnamefont {M.~G.}\ \bibnamefont {Mingolla}}, \bibinfo {author}
  {\bibfnamefont {F.}~\bibnamefont {Piacentini}}, \bibinfo {author}
  {\bibfnamefont {M.}~\bibnamefont {Rajteri}}, \bibinfo {author} {\bibfnamefont
  {E.}~\bibnamefont {Taralli}}, \ and\ \bibinfo {author} {\bibfnamefont
  {M.~G.~A.}\ \bibnamefont {Paris}},\ }\bibfield  {title} {\enquote {\bibinfo
  {title} {Quantum characterization of superconducting photon counters},}\
  }\href@noop {} {\bibfield  {journal} {\bibinfo  {journal} {New J. Phys.}\
  }\textbf {\bibinfo {volume} {14}},\ \bibinfo {pages} {085001} (\bibinfo
  {year} {2012})}\BibitemShut {NoStop}%
\bibitem [{\citenamefont {{\v R}eh{\'a}{\v c}ek}\ \emph
  {et~al.}(2010)\citenamefont {{\v R}eh{\'a}{\v c}ek}, \citenamefont
  {Mogilevtsev},\ and\ \citenamefont {Hradil}}]{Rehacek:2010fk}%
  \BibitemOpen
  \bibfield  {author} {\bibinfo {author} {\bibfnamefont {J.}~\bibnamefont {{\v
  R}eh{\'a}{\v c}ek}}, \bibinfo {author} {\bibfnamefont {D.}~\bibnamefont
  {Mogilevtsev}}, \ and\ \bibinfo {author} {\bibfnamefont {Z.}~\bibnamefont
  {Hradil}},\ }\bibfield  {title} {\enquote {\bibinfo {title} {Operational
  tomography: Fitting of data patterns},}\ }\href@noop {} {\bibfield  {journal}
  {\bibinfo  {journal} {Phys. Rev. Lett.}\ }\textbf {\bibinfo {volume} {105}},\
  \bibinfo {pages} {01040} (\bibinfo {year} {2010})}\BibitemShut {NoStop}%
\bibitem [{\citenamefont {Mogilevtsev}\ \emph {et~al.}(2013)\citenamefont
  {Mogilevtsev}, \citenamefont {Ignatenko}, \citenamefont {Maloshtan},
  \citenamefont {Stoklasa}, \citenamefont {Rehacek},\ and\ \citenamefont
  {Hradil}}]{Mogilevtsev:2013kl}%
  \BibitemOpen
  \bibfield  {author} {\bibinfo {author} {\bibfnamefont {D.}~\bibnamefont
  {Mogilevtsev}}, \bibinfo {author} {\bibfnamefont {A.}~\bibnamefont
  {Ignatenko}}, \bibinfo {author} {\bibfnamefont {A.}~\bibnamefont
  {Maloshtan}}, \bibinfo {author} {\bibfnamefont {B.}~\bibnamefont {Stoklasa}},
  \bibinfo {author} {\bibfnamefont {J.}~\bibnamefont {Rehacek}}, \ and\
  \bibinfo {author} {\bibfnamefont {Z.}~\bibnamefont {Hradil}},\ }\bibfield
  {title} {\enquote {\bibinfo {title} {Data pattern tomography: reconstruction
  with an unknown apparatus},}\ }\href@noop {} {\bibfield  {journal} {\bibinfo
  {journal} {New J. Phys.}\ }\textbf {\bibinfo {volume} {15}},\ \bibinfo
  {pages} {025038} (\bibinfo {year} {2013})}\BibitemShut {NoStop}%
\bibitem [{\citenamefont {Cooper}\ \emph {et~al.}(2014)\citenamefont {Cooper},
  \citenamefont {Karpinski},\ and\ \citenamefont {Smith}}]{Cooper:2014qf}%
  \BibitemOpen
  \bibfield  {author} {\bibinfo {author} {\bibfnamefont {M.}~\bibnamefont
  {Cooper}}, \bibinfo {author} {\bibfnamefont {M.}~\bibnamefont {Karpinski}}, \
  and\ \bibinfo {author} {\bibfnamefont {B.~J.}\ \bibnamefont {Smith}},\
  }\bibfield  {title} {\enquote {\bibinfo {title} {Local mapping of detector
  response for reliable quantum state estimation},}\ }\href@noop {} {\bibfield
  {journal} {\bibinfo  {journal} {Nat. Commun.}\ }\textbf {\bibinfo {volume}
  {5}},\ \bibinfo {pages} {4332} (\bibinfo {year} {2014})}\BibitemShut
  {NoStop}%
\bibitem [{\citenamefont {Harder}\ \emph {et~al.}(2014)\citenamefont {Harder},
  \citenamefont {Silberhorn}, \citenamefont {Rehacek}, \citenamefont {Hradil},
  \citenamefont {Motka}, \citenamefont {Stoklasa},\ and\ \citenamefont
  {S{\'a}nchez-Soto}}]{Harder:2014tf}%
  \BibitemOpen
  \bibfield  {author} {\bibinfo {author} {\bibfnamefont {G.}~\bibnamefont
  {Harder}}, \bibinfo {author} {\bibfnamefont {C.}~\bibnamefont {Silberhorn}},
  \bibinfo {author} {\bibfnamefont {J.}~\bibnamefont {Rehacek}}, \bibinfo
  {author} {\bibfnamefont {Z.}~\bibnamefont {Hradil}}, \bibinfo {author}
  {\bibfnamefont {L.}~\bibnamefont {Motka}}, \bibinfo {author} {\bibfnamefont
  {B.}~\bibnamefont {Stoklasa}}, \ and\ \bibinfo {author} {\bibfnamefont
  {L.~L.}\ \bibnamefont {S{\'a}nchez-Soto}},\ }\bibfield  {title} {\enquote
  {\bibinfo {title} {Time-multiplexed measurements of nonclassical light at
  telecom wavelengths},}\ }\href@noop {} {\bibfield  {journal} {\bibinfo
  {journal} {Phys. Rev. A}\ }\textbf {\bibinfo {volume} {90}},\ \bibinfo
  {pages} {042105} (\bibinfo {year} {2014})}\BibitemShut {NoStop}%
\bibitem [{\citenamefont {Altorio}\ \emph {et~al.}(2016)\citenamefont
  {Altorio}, \citenamefont {Genoni}, \citenamefont {Somma},\ and\ \citenamefont
  {Barbieri}}]{Altorio:2016aa}%
  \BibitemOpen
  \bibfield  {author} {\bibinfo {author} {\bibfnamefont {M.}~\bibnamefont
  {Altorio}}, \bibinfo {author} {\bibfnamefont {M.~G.}\ \bibnamefont {Genoni}},
  \bibinfo {author} {\bibfnamefont {F.}~\bibnamefont {Somma}}, \ and\ \bibinfo
  {author} {\bibfnamefont {M.}~\bibnamefont {Barbieri}},\ }\bibfield  {title}
  {\enquote {\bibinfo {title} {Metrology with unknown detectors},}\ }\href@noop
  {} {\bibfield  {journal} {\bibinfo  {journal} {Phys. Rev. Lett.}\ }\textbf
  {\bibinfo {volume} {116}},\ \bibinfo {pages} {100802} (\bibinfo {year}
  {2016})}\BibitemShut {NoStop}%
\bibitem [{\citenamefont {Harder}\ \emph {et~al.}(2013)\citenamefont {Harder},
  \citenamefont {Ansari}, \citenamefont {Brecht}, \citenamefont {Dirmeier},
  \citenamefont {Marquardt},\ and\ \citenamefont {Silberhorn}}]{Harder:2013iq}%
  \BibitemOpen
  \bibfield  {author} {\bibinfo {author} {\bibfnamefont {G.}~\bibnamefont
  {Harder}}, \bibinfo {author} {\bibfnamefont {V.}~\bibnamefont {Ansari}},
  \bibinfo {author} {\bibfnamefont {B.}~\bibnamefont {Brecht}}, \bibinfo
  {author} {\bibfnamefont {T.}~\bibnamefont {Dirmeier}}, \bibinfo {author}
  {\bibfnamefont {C.}~\bibnamefont {Marquardt}}, \ and\ \bibinfo {author}
  {\bibfnamefont {C.}~\bibnamefont {Silberhorn}},\ }\bibfield  {title}
  {\enquote {\bibinfo {title} {An optimized photon pair source for quantum
  circuits},}\ }\href@noop {} {\bibfield  {journal} {\bibinfo  {journal} {Opt.
  Express}\ }\textbf {\bibinfo {volume} {12}},\ \bibinfo {pages} {13975--13985}
  (\bibinfo {year} {2013})}\BibitemShut {NoStop}%
\bibitem [{sup()}]{suppl}%
  \BibitemOpen
  \href@noop {} {\enquote {\bibinfo {title} {See supplemental material at},}\
  }\BibitemShut {NoStop}%
\end{thebibliography}
%

\end{document}